\documentclass[twocolumn,floatfix,aps,superscriptaddress,eqsecnum,prb.1]{revtex4}
\usepackage{graphicx}
\usepackage{amsmath}
\usepackage{amssymb}
\usepackage{bm}

\def\Xint#1{\mathchoice
   {\XXint\displaystyle\textstyle{#1}}%
   {\XXint\textstyle\scriptstyle{#1}}%
   {\XXint\scriptstyle\scriptscriptstyle{#1}}%
   {\XXint\scriptscriptstyle\scriptscriptstyle{#1}}%
   \!\int}
\def\XXint#1#2#3{{\setbox0=\hbox{$#1{#2#3}{\int}$}
     \vcenter{\hbox{$#2#3$}}\kern-.5\wd0}}

\def\dashint{\Xint-}

\usepackage{color}

\begin{document}

\title{Half-integer charge injection by a Josephson junction without excess noise}
\author{F. Hassler}
\affiliation{JARA-Institute for Quantum Information, RWTH Aachen University, 52056 Aachen, Germany}
\author{A. Grabsch}
\affiliation{Instituut-Lorentz, Universiteit Leiden, P.O. Box 9506, 2300 RA Leiden, The Netherlands}
\author{M. J. Pacholski}
\affiliation{Instituut-Lorentz, Universiteit Leiden, P.O. Box 9506, 2300 RA Leiden, The Netherlands}
\author{D. O. Oriekhov}
\affiliation{Instituut-Lorentz, Universiteit Leiden, P.O. Box 9506, 2300 RA Leiden, The Netherlands}
\affiliation{Department of Physics, Taras Shevchenko National University of Kyiv, Kyiv 03680, Ukraine}
\author{O. Ovdat}
\affiliation{Instituut-Lorentz, Universiteit Leiden, P.O. Box 9506, 2300 RA Leiden, The Netherlands}
\author{I. Adagideli}
\affiliation{Faculty of Engineering and Natural Sciences,
Sabanc{\i} University, Orhanl{\i}-Tuzla, 34956, Turkey}
\affiliation{Faculty of Science and Technology and MESA+ Institute for Nanotechnology, University of Twente, 7500 AE Enschede, The Netherlands}
\author{C. W. J. Beenakker}
\affiliation{Instituut-Lorentz, Universiteit Leiden, P.O. Box 9506, 2300 RA Leiden, The Netherlands}

\date{May 2020}
\begin{abstract}
A Josephson junction in a topological superconductor can inject a charge $e/2$ into a normal-metal contact, carried by chiral Majorana edge modes. Here we address the question whether this half-integer charge is a sharp observable, without quantum fluctuations. Because the Majorana modes are gapless, they support charge fluctuations in equilibrium at zero temperature. But we find that the excess noise introduced out of equilibrium by the $e/2$ charge transfer vanishes. We discuss a strategy to reduce the equilibrium fluctuations, by means of a heavy-tailed time-dependent detection efficiency, to achieve a nearly noiseless half-integer charge transfer.
\end{abstract}
\maketitle

\section{Introduction}
\label{intro}

Chiral Majorana edge modes in a topological superconductor can be excited by a flux bias applied to a Josephson junction \cite{Bee19a}, analogously to the excitation of chiral Dirac edge modes in a quantum Hall insulator by a voltage pulse applied to a tunnel junction \cite{Gre11}. There is a key difference: The elementary excitation of a Dirac edge mode has charge $e$, produced by a $2\pi$ phase increment of the single-electron wave function \cite{Lev96,Iva97,Kee06,Dub13}. In a superconductor, in contrast, a $2\pi$ phase increment of the pair potential is a $\pi$ phase shift for single electrons. This explains why an $h/2e$ flux bias of a Josephson junction transfers charge $e/2$ into a normal metal contact \cite{Ada20}.

\begin{figure}[tb]
\centerline{\includegraphics[width=0.9\linewidth]{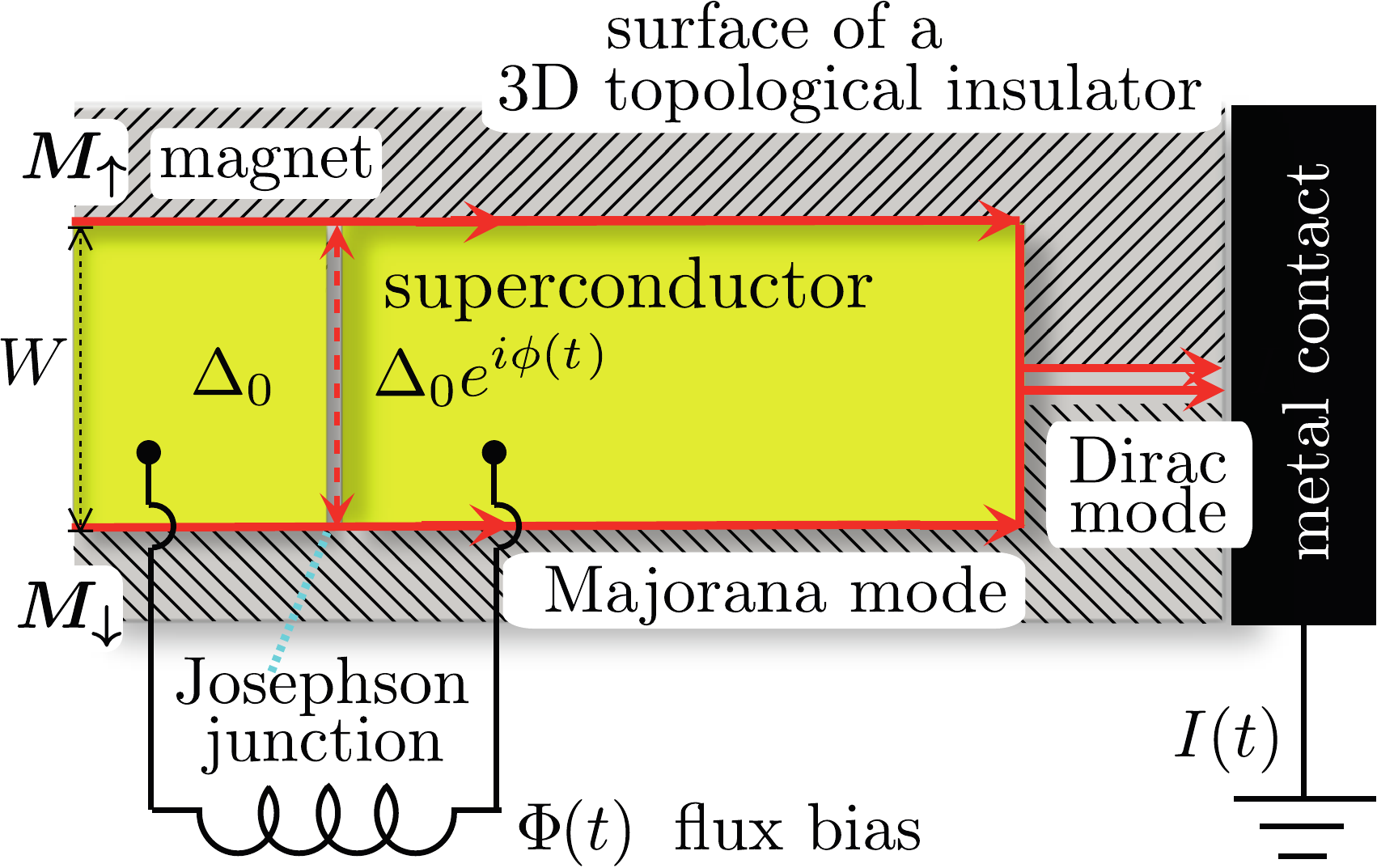}}
\caption{Geometry to inject a half-integer charge in a topological insulator/magnetic insulator/superconductor heterostructure \cite{Bee19a,Ada20}. An $h/2e$ flux bias $\Phi(t)$ induces a $2\pi$ increment of the superconducting phase difference $\phi(t)$ across a Josephson junction. A pair of edge vortices is excited in the Majorana edge modes of the superconductor, which fuse to form a Dirac mode when they enter a normal-metal contact. The resulting current pulse $I(t)$ has integrated charge $Q=\pm e/2$. The Dirac mode in this diagram propagates in a single direction only. The counterpropagating mode does not couple to the superconductor, so it can be ignored.
}
\label{fig_layout}
\end{figure}

This half-integer charge is encoded nonlocally in a pair of $\pi$-phase domain walls, or ``edge vortices'' \cite{Fen07}, which propagate away from the Josephson junction along the Majorana modes at opposite edges (see Fig.\ \ref{fig_layout}). When the Majorana modes reach the metal contact they are merged into a Dirac mode and the charge can be detected \cite{Fu09,Akh09}. Since only integer charge can enter into a normal metal, the electrical current cannot be completely noiseless --- as it can be in the single-electron case \cite{Lev96,Iva97,Kee06,Dub13}. What is the noise associated with the fractional charge transfer? That is the question addressed in this paper.

The $e/2$ charge carried by Majorana edge modes can be seen as the mobile counterpart of the $e/2$ charge bound to a zero-mode in a topological insulator \cite{Jac76,Su79,Nie86,Hou07,Ben19}. In that context it is known that the half-integer charge is a sharp observable \cite{Kiv82,Bel82,Jac83}, without quantum fluctuations, provided that the charge of the zero-mode is measured by integrating the charge density over a wide region with smooth boundaries --- in order to work around the integer charge constraint. The analogy between mobile edge modes and immobile zero-modes is suggestive, but limited: The zero-mode is embedded in an excitation gap, while the Majorana edge modes are gapless. We can therefore expect charge fluctuations to play a more significant role in the latter case.

The outline of the paper is as follows. After a brief description of the physical system in the next section we formulate the mathematical problem in Sec.\ \ref{sec_cumulant}. The main result of that section is an expression for the cumulant generating function in terms of ``anchored'' scattering matrices --- meaning that the expression is dominated by scattering processes near the Fermi level. Anchoring to the Fermi level is a crucial step when one models the chiral modes by a Fermi sea with an unbounded linear dispersion. 

The second cumulant (the charge variance) is calculated in Sec.\ \ref{sec_noise} and then all higher cumulants are obtained in Sec. \ref{sec_higher}. Because the edge modes are gapless, there is charge noise even in equilibrium at zero temperature \cite{Lev96}. However, what we find is that the \textit{excess noise} produced out of equilibrium by the $e/2$ charge transfer vanishes. In Sec.\ \ref{sec_reduce} we propose a strategy to reduce the equilibrium noise by smoothing the detection window, so that the fractional charge can truly become a sharp observable. We conclude in Sec.\ \ref{sec_conclude} with a proposal for an experiment and by making the connection with fractional charge transfer in normal metals \cite{Mos16,Yue20}.

\section{Charge injection by a Josephson junction}

To set the stage, we summarize results from Refs.\ \onlinecite{Bee19a,Ada20} for the Josephson junction geometry of Fig.\ \ref{fig_layout}. The junction connects one-dimensional Majorana modes propagating unidirectionally (chirally) along opposite edges of a two-dimensional topological superconductor. The Majorana modes are excited by a $2\pi$ increment of the phase difference $\phi(t)$ across the Josephson junction. While the pair potential $\Delta=\Delta_0 e^{i\phi}$ returns to its original value after the phase jump, the same does not hold for the edge modes: The two Majorana modes together form an electronic degree of freedom for which a $2\pi$ increment of the superconducting phase corresponds to a phase jump of $\pi$. The $\pi$-phase boundary propagates away from the junction along the edge modes as a pair of edge vortices \cite{Fen07}, one on each edge.

While each Majorana mode separately is charge neutral, a nonzero charge may appear when the Majorana modes are fused into a single Dirac mode at a metal contact \cite{Fu09,Akh09}. The charge transferred by the fused edge vortices depends on the relative magnitude of the path length difference $\delta L$ along the upper and lower edge and the product $vt_{\rm inj}$ of the edge mode velocity $v$ and the duration $t_{\rm inj}$ of the vortex injection process --- a time scale given by \cite{Bee19a}
\begin{equation}
t_{\rm inj}= (\xi_0/W)(d\phi/dt)^{-1},\label{tphidef}
\end{equation}
with $\xi_0=\hbar v/\Delta_0$ the superconducting coherence length and $W$ the width of the Josephson junction. For $W/\xi_0\gg 1$ the fusion of the edge vortices injects an average charge 
\begin{equation}
\langle Q\rangle=\pm\frac{e}{2}\frac{\tanh(\delta L/4vt_{\rm inj})}{\delta L/4vt_{\rm inj}}\label{Qtdelay}
\end{equation}
into the metal contact \cite{Ada20}. (The sign depends on the sign of the phase change.) A maximal charge of $\pm e/2$ is transferred for $\delta L\ll vt_{\rm inj}$.

This half-integer charge transfer is the average over many measurements. One might intuitively expect a binomial statistics, where charge $0$ or $e$ is transferred with equal probability. That would produce a charge variance of $e^2/4$, on top of any equilibrium noise. As we will see in the following section, that intuitive expectation is completely mistaken, and in fact the charge transfer for $\delta L=0$ introduces no excess noise at all.

\section{Cumulant generating function}
\label{sec_cumulant}

\subsection{Determinantal formula}

The generating function $C(\xi)$ of cumulants of transferred charge is given by
\begin{equation}
C(\xi)=\ln\,{\rm Tr}\,\bigl(\hat{\rho}_{\rm eq} e^{i\xi \hat{Q}}\bigr),\label{Fxidef}
\end{equation}
in terms of the charge operator $\hat{Q}$ of the outgoing modes and the equilibrium density matrix $\hat{\rho}_{\rm eq}$ of the incoming modes. (For ease of notation we set the electron charge $e$ to unity, as well as the constants $\hbar$ and $k_{\rm B}$, restoring these in final equations.)

The incoming Majorana modes have annihilation operators $\hat{a}_n(E)$, with $n=1,2$ for the upper and lower edge. Positive and negative energies are related by particle-hole symmetry,
\begin{equation}
\hat{a}_n(-E)=\hat{a}^\dagger_n(E).\label{aadaggerphsym}
\end{equation}
Similarly, $\hat{b}_n(E)$ is the annihilation operator for the outgoing modes. Incoming and outgoing modes are related by the scattering matrix, 
\begin{equation}
\hat{b}_n(E)=\int_{-\infty}^\infty \frac{dE'}{2\pi}\,\textstyle{\sum_{m}}S_{nm}(E,E')\hat{a}_m(E').
\end{equation}
The scattering matrix is unitary,
\begin{align}
\int_{-\infty}^\infty \frac{dE}{2\pi}\textstyle{\sum_{n'}}S_{n'n}^\ast(E,E_1)S_{n'm}(E,E_2)=\nonumber\\
=2\pi\delta(E_1-E_2)\delta_{nm},
\end{align}
and it satisfies particle-hole symmetry,
\begin{equation}
S_{nm}(-E,-E')=S^\ast_{nm}(E,E').
\end{equation}

The charge operator for Majorana modes is
\begin{equation}
\hat{Q}=\int_{0}^\infty\frac{dE}{2\pi}\,\bigl(i\hat{b}_2^\dagger(E)\hat{b}_1(E)-i\hat{b}_1^\dagger(E)\hat{b}_2(E)\bigr).
\end{equation}
The particle-hole symmetry relation \eqref{aadaggerphsym} allows us to extend the integration range $\int_0^\infty dE\mapsto \tfrac{1}{2}\int_{-\infty}^\infty dE$. We then write more compactly
\begin{equation}
\hat{Q}=\tfrac{1}{2}\hat{b}^\dagger\cdot\sigma_y\cdot\hat{b}=\tfrac{1}{2}\hat{a}^\dagger \cdot S^\dagger\sigma_y S\cdot\hat{a},\label{Qdef}
\end{equation}
with $\sigma_y$ a Pauli matrix acting on the mode indices. The contraction $\cdot$ indicates both a sum over the mode index and an integration over energy. The kernel $S^\dagger\sigma_y S$ satisfies a generalized antisymmetry relation,
\begin{equation}
[S^\dagger\sigma_y S]_{nm}(E,E')=-[S^\dagger\sigma_y S]_{mn}(-E',-E).\label{Santisymmetry}
\end{equation}

We now use the Klich formula \cite{Kli14,Bee19b} to reduce the operator trace \eqref{Fxidef} to a functional determinant,
\begin{equation}
{\rm Tr}\,\bigl(\hat{\rho}_{\rm eq} e^{\hat{a}^\dagger\cdot M\cdot \hat{a}}\bigr)=\sqrt{e^{{\rm Tr}\,M}\,{\rm Det}\,\bigl(1-{\cal F}+{\cal F} e^{2M^{\rm A}}\bigr)},\label{Klichformula}
\end{equation}
with the definitions
\begin{align}
&M^{\rm A}_{nm}(E,E')=\tfrac{1}{2}M_{nm}(E,E')-\tfrac{1}{2}M_{mn}(-E,-E'),\nonumber\\
&{\cal F}(E,E')=2\pi\delta(E-E')f(E).
\end{align}
The Fermi function
\begin{equation}
f(E)=(1+e^{E/T})^{-1}
\end{equation}
is the occupation number of the $\hat{a}$ modes, in equilibrium at temperature $T$ and Fermi energy $E_{\rm F}=0$.

In view of Eqs.\ \eqref{Qdef} and \eqref{Santisymmetry} the cumulant generating function takes the form
\begin{equation}
C(\xi)=\tfrac{1}{2}\ln\,{\rm Det}\,\bigl(1-{\cal F}+{\cal F}S^\dagger e^{i\xi\sigma_y}S \bigr).\label{Fxidet}
\end{equation}
This is the expression we seek to evaluate.

\subsection{Anchored scattering matrix}

Functional determinants of the type \eqref{Fxidet} need to be regularized before they can be applied to an unbounded spectrum \cite{Iva97,Muz03,Aba08,Avr08,note1}.  For that purpose we rewrite the determinant such that the scattering matrix contributes only for energies near the Fermi level ($E=0$). We ``anchor'' the scattering matrix to the Fermi level by commutating it with the Fermi function,
\begin{equation}
\begin{split}
&\tilde{S}={\cal F}S-S{\cal F},\\
&\tilde{S}_{nm}(E,E')=S_{nm}(E,E')[f(E)-f(E')].
\end{split}\label{Sanchored}
\end{equation}
The kernel $\tilde{S}(E,E')$ vanishes unless $E,E'$ are within $\max(T,1/t_{\rm inj})$ from the Fermi level. 

Substitution into Eq.\ \eqref{Fxidet} gives
\begin{equation}
C(\xi)=\tfrac{1}{2}\ln\,{\rm Det}\,\bigl[1+\bigl(e^{i\xi\sigma_y}-1\bigr)({\cal F}-\tilde{S}S^\dagger )\bigr].\label{Fxidet2}
\end{equation}
In equilibrium the scattering is elastic, hence $\tilde{S}$ vanishes. It is useful to extract from Eq.\ \eqref{Fxidet2} the equilibrium generating function by decomposing $C=C_{\rm eq}+\delta C$, with
\begin{subequations}
\label{Fxidet3}
\begin{align}
&C_{\rm eq}(\xi)=\tfrac{1}{2}\ln\,{\rm Det}\,\bigl(1+\bigl(e^{i\xi\sigma_y}-1\bigr){\cal F}\bigr),\\
&\delta C(\xi)=\tfrac{1}{2}\ln\,{\rm Det}\,(1-\Xi(\xi)\tilde{S}S^\dagger),\\&\Xi(\xi)=\frac{e^{i\xi\sigma_y}-1}{1+\bigl(e^{i\xi\sigma_y}-1\bigr){\cal F}}.
\end{align}
\end{subequations}

\section{Calculation of the excess noise}
\label{sec_noise}

\subsection{Expansion of the cumulant generating function}

The variance $\langle\!\langle Q^2\rangle\!\rangle=\langle Q^2\rangle-\langle Q\rangle^2$ of the transferred charge contains an equilibrium contribution $\langle\!\langle Q^2\rangle\!\rangle_{\rm eq}$ from $C_{\rm eq}(\xi)$ plus a contribution $\delta\langle\!\langle Q^2\rangle\!\rangle$ from $\delta C(\xi)$. The latter contribution is the excess charge noise introduced by the vortex injection process.

We calculate both contributions, as well as the average transferred charge $\langle Q\rangle$, by expanding the cumulant generating function \eqref{Fxidet3} to second order in $\xi$,
\begin{equation}
C(\xi)=i\xi\langle Q\rangle-\tfrac{1}{2}\xi^2\langle\!\langle Q^2\rangle\!\rangle+{\cal O}(\xi^3).
\end{equation}
The expansion is carried out by means of the identity $\ln \,{\rm Det}\,M={\rm Tr}\,\ln M$ and Taylor expansion of the logarithm. This results in
\begin{align}
&\langle Q\rangle=-\tfrac{1}{2}\,{\rm Tr}\,\sigma_y\tilde{S}S^\dagger,\label{firsttwocumulantsa}
\\
&\langle\!\langle Q^2\rangle\!\rangle_{\rm eq}=\tfrac{1}{2}\,{\rm Tr}\,\sigma_0{\cal F}{\cal F}_c,\label{firsttwocumulantsb}\\
&\delta\langle\!\langle Q^2\rangle\!\rangle=\tfrac{1}{2}\,{\rm Tr}\,\bigl[({\cal F}-{\cal F}_c)\tilde{S}S^\dagger-(\sigma_y\tilde{S}S^\dagger)^2\bigr],\label{firsttwocumulantsc}
\end{align}
with $\sigma_0$ the $2\times 2$ unit matrix and ${\cal F}_c=1-{\cal F}$.

\subsection{Equilibrium noise}

The charge noise in equilibrium has been obtained by Levitov, Lee, and Lesovik \cite{Lev96}. We calculate it here for later reference, because we will need some of the equations for the calculation of the nonequilibrium noise.

The Fermi function in the time domain is given by
\begin{align}
{\cal F}(t,t')&=\int_{-\infty}^{\infty}\frac{dE}{2\pi}e^{-iE(t-t'+ i\epsilon)}\frac{1}{1+e^{ E/T}}\nonumber\\
&=\frac{iT}{2\sinh[\pi T(t-t'+ i\epsilon)]}={\cal F}_c(t',t),
\end{align}
with $\epsilon$ a positive infinitesimal set by the inverse band width. For later use we also note that
\begin{equation}
[{\cal F}-{\cal F}_c](t,t')={\cal P}\frac{iT}{\sinh[\pi T(t-t')]},\label{barFminF}
\end{equation}
where ${\cal P}$ indicates the Cauchy principal value.

From Eq.\ \eqref{firsttwocumulantsb}, the equilibrium charge variance is given by
\begin{equation}
\langle\!\langle Q^2\rangle\!\rangle_{\rm eq}=\int_{-t_{\rm det}/2}^{t_{\rm det}/2}dt\int_{-t_{\rm det}/2}^{t_{\rm det}/2}dt'\,{\cal F}(t,t'){\cal F}_c(t',t),\label{varQeqFttprime}
\end{equation}
where we have introduced a finite detection time $t_{\rm det}$. We thus recover the result of Ref.\ \onlinecite{Lev96},
\begin{align}
\langle\!\langle Q^2\rangle\!\rangle_{\rm eq}&=\int_{-t_{\rm det}/2}^{t_{\rm det}/2}dt\int_{-t_{\rm det}/2}^{t_{\rm det}/2}dt'\,\frac{-T^2}{4\sinh^2[\pi T(t-t'+i\epsilon)]}\nonumber\\
&=\frac{1}{2\pi^2}\ln\left(\frac{\sinh(\pi Tt_{\rm det})}{\pi T\epsilon}\right)+{\cal O}(\epsilon/t_{\rm det})^2.\label{eqnoisefinal}
\end{align}

The equilibrium noise increases linearly $\propto Tt_{\rm det}$ with the detection time for $Tt_{\rm det}\gg 1$, while it increases logarithmically $\propto\ln(t_{\rm det}/\epsilon)$ for $Tt_{\rm det}\ll 1$. If we restore units of $e$, $k_{\rm B}$, and $\hbar$, the equilibrium noise corresponds to a noise power 
\begin{equation}
P_{\rm eq}=\lim_{\rm t_{\rm det}\rightarrow\infty}\frac{1}{t_{\rm det}}\langle\!\langle Q^2\rangle\!\rangle_{\rm eq}=\frac{e^2}{2\pi\hbar}k_{\rm B}T,
\end{equation}
which is one-half the Johnson-Nyquist noise for a single-mode conductor --- the other half comes from a counterpropagating Dirac mode that does not couple to the superconductor (see Fig.\ \ref{fig_layout}).

\subsection{Adiabatic scattering matrix}

In the time domain, upon Fourier transformation according to
\begin{equation}
S(t,t')=\int_{-\infty}^\infty\frac{dE}{2\pi}\int_{-\infty}^\infty\frac{dE'}{2\pi}\,e^{iE't'-iEt}S(E,E'),
\end{equation}
the scattering matrix $S(t,t')$ becomes a delta function in the adiabatic approximation,
\begin{equation}
S(t,t')=S_0(t)\delta(t-t')+{\cal O}(t_{\rm dwell}/t_{\rm inj}),\label{Sadiabatic}
\end{equation}
valid if the dwell time $t_{\rm dwell}\simeq W/v$ in the Josephson junction is small compared to the vortex injection time $t_{\rm inj}$. 

The $2\times 2$ unitary matrix $S_0(t)$ is the socalled ``frozen'' scattering matrix of the Josephson junction, obtained by fixing the phase at its instantaneous value of $\phi(t)$. It has the form \cite{Bee19a}
\begin{equation}
S_0(t)=e^{i\eta(t)\sigma_y},\label{S0eta}
\end{equation}
with a phase $\eta(t)$ that increases by $\pi$ on a time scale $t_{\rm inj}$. The profile by which $\eta(t)$ increases depends on the width $W$ of the junction relative to the superconducting coherence length $\xi_0$. For $W\gg\xi_0$ one has
\begin{equation}
\eta(t)= \arccos[-\tanh(t/2t_{\rm inj})].\label{etawidejunction}
\end{equation}

There may be a relative delay $\delta t=\delta L/v$ in the propagation time along the upper or lower edge from Josephson junction to metal contact. This can be included via the replacement in Eq.\ \eqref{Sadiabatic} of the single delta function by a separate delta function for each Majorana mode,
\begin{equation}
S(t,t')=\begin{pmatrix}
\delta(t-t')&0\\
0&\delta(t-t'+\delta t)
\end{pmatrix}S_0(t').\label{Swithdelay}
\end{equation}
In what follows we will address the case $\delta t=0$ in the main text, turning to the effect of a nonzero relative delay in App.\ \ref{app_delay}.

The adiabatic scattering matrix $S(t,t')=S_0(t)\delta(t-t')$ is singular at $t=t'$. Anchoring via the commutator $\tilde S=[{\cal F},S]$ removes the singularity,
\begin{equation}
\tilde{S}(t,t')=\frac{T}{2i}\frac{S_0(t)-S_0(t')}{\sinh[\pi T (t-t')]},\;\;\tilde{S}(t,t)=\frac{1}{2\pi i}\frac{d}{dt}S_0(t).\label{tildeSresult}
\end{equation}
(No $\epsilon$-regularization is needed for a nonsingular kernel.) 

Eq.\ \eqref{firsttwocumulantsa} now immediately reproduces the superconducting analogue of Brouwer's charge-pumping formula \cite{Bro98,Tar15},
\begin{align}
\langle Q\rangle=\frac{ie}{4\pi}\,\int_{-t_{\rm det}/2}^{t_{\rm det}/2}dt\,{\rm Tr}\, S_0^\dagger(t)\sigma_y \frac{d}{dt}S_0(t).
\end{align}
Substitution of Eq.\ \eqref{S0eta} then gives the half-integer average charge transfer \cite{Bee19a,Ada20},
\begin{equation}
\langle Q\rangle=-\frac{e}{2\pi}\int_{-t_{\rm det}/2}^{t_{\rm det}/2}dt\,\eta'(t)=-e/2,\;\;t_{\rm det}\gg t_{\rm inj},\label{Qaveragefinal}
\end{equation}
for any phase profile $\eta(t)$ that winds by $\pi$, independent of temperature for $Tt_{\rm dwell}\ll 1$.

\subsection{Excess noise}

Turning now to the nonequilibrium contribution $\delta\langle\!\langle Q^2\rangle\!\rangle$  to the variance, we have upon substitution of Eqs.\ \eqref{barFminF} and \eqref{tildeSresult} into Eq.\ \eqref{firsttwocumulantsc},
\begin{widetext}
\begin{align}
\delta\langle\!\langle Q^2\rangle\!\rangle={}&T^2\int_{-t_{\rm det}/2}^{t_{\rm det}/2}dt\int_{-t_{\rm det}/2}^{t_{\rm det}/2}dt'\,\frac{{\rm Tr}\,S^\dagger(t)[S(t)-S(t')]}{4\sinh^2[\pi T(t-t')]}\nonumber\\
&-T^2\int_{-t_{\rm det}/2}^{t_{\rm det}/2}dt\int_{-t_{\rm det}/2}^{t_{\rm det}/2}dt'\,\frac{{\rm Tr}\,\sigma_y[S(t)-S(t')]S^\dagger(t')\sigma_y[S(t')-S(t)]S^\dagger(t)}{8\sinh^2[\pi T(t-t')]}.\label{deltaQintegral}
\end{align}

Upon substitution of $S(t)=e^{i\eta(t)\sigma_y}$ we find that
\begin{align}
2\,{\rm Tr}\,S^\dagger(t)[S(t)-S(t')]&={\rm Tr}\,\sigma_y[S(t)-S(t')]S^\dagger(t')\sigma_y[S(t')-S(t)]S^\dagger(t)=8\sin^2[\tfrac{1}{2}\eta(t)-\tfrac{1}{2}\eta(t')],\label{TrSdagSidentity}
\end{align}
\end{widetext}
hence both integrands in Eq.\ \eqref{deltaQintegral} are nonsingular at $t=t'$ (no principal value is needed) and moreover they cancel. We conclude that
\begin{equation}
\delta\langle\!\langle Q^2\rangle\!\rangle=0,
\end{equation}
the excess charge noise vanishes \textit{at any temperature}, only the equilibrium noise \eqref{eqnoisefinal} remains.

\begin{figure}[tb]
\centerline{\includegraphics[width=0.9\linewidth]{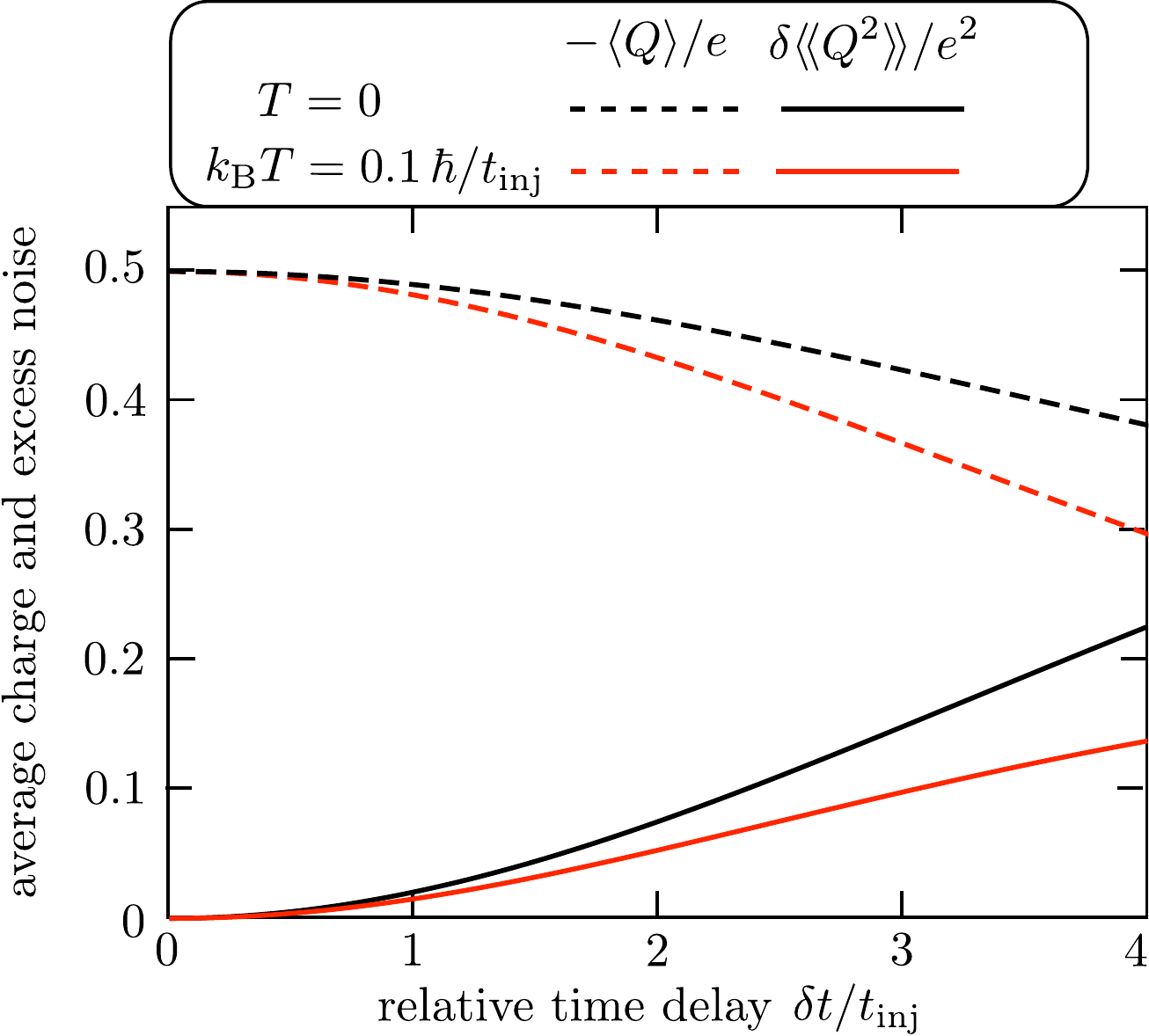}}
\caption{Dependence on the relative time delay $\delta t$ of the two Majorana modes of the average transferred charge (dashed curves) and the excess noise (solid curves), at zero temperature (black curves) and at nonzero temperature (red curves). The curves are calculated for the phase profile \eqref{etawidejunction}, valid in the limit $W\gg\xi_0$ of a wide junction, from the expressions \eqref{Qdeltattemp} and \eqref{deltaQtemp}. The $T=0$ result for $\langle Q\rangle$ is also given by Eq.\ \eqref{Qtdelay}.
}
\label{fig_noise}
\end{figure}

All of this is for zero relative delay between the two Majorana modes. The case of nonzero $\delta t$ is calculated in App.\ \ref{app_delay}, results are shown in Fig.\ \ref{fig_noise}. When the delay time becomes comparable to the injection time $t_{\rm inj}$ the average transferred charge is reduced below $e/2$, and a nonzero excess noise appears. The excess noise decreases with temperature, presumably because of thermal averaging of a state which is not an eigenstate of charge. 

\section{Calculation of higher cumulants}
\label{sec_higher}

If we restrict ourselves to zero temperature and zero time delay, we can go beyond mean and variance and calculate all cumulants of the transferred charge.

\subsection{Even cumulants}

We start from Eq.\ \eqref{Fxidet3} and take the derivative with respect to $\xi$ of the nonequilibrium cumulant generating function,
\begin{align}
\frac{d}{d\xi}\delta C(\xi)={}&\frac{1}{2}\frac{d}{d\xi}\,{\rm Tr}\,\ln(1-\Xi(\xi)\tilde{S}S^\dagger)\nonumber\\
={}&-\tfrac{1}{2}\,{\rm Tr}\,\bigl[(1-\Xi(\xi)\tilde{S}S^\dagger)^{-1}\Xi'(\xi)\tilde{S}S^\dagger\bigr].\label{ddxideltaFxi}
\end{align}
We evaluate $\Xi(\xi)$ at $T=0$, when ${\cal F}^2={\cal F}$, hence
\begin{align}
\Xi(\xi)&=(e^{i\xi\sigma_y}-1)(1-{\cal F})+(1-e^{-i\xi\sigma_y}){\cal F}\nonumber\\
&=(1-\cos\xi)({\cal F}-{\cal F}_c)+i\sigma_y\sin\xi.
\end{align}
Recall that ${\cal F}_c=1-{\cal F}$. 

In App.\ \ref{app_identitiesA} we derive that
\begin{equation}
({\cal F}-{\cal F}_c)\tilde{S}S^\dagger+\tilde{S}S^\dagger({\cal F}-{\cal F}_c)=2(\tilde{S}S^\dagger)^2.\label{keyidentity}
\end{equation}
From this identity, and from $({\cal F}-{\cal F}_c)^2=1$, it follows that
\begin{equation}
\bigl(1-\tilde{S}S^\dagger\Xi(-\xi)\bigr)\bigl(1-\Xi(\xi)\tilde{S}S^\dagger\bigr)=1,
\end{equation}
so these two operators are each others inverse. We can then work out the derivative \eqref{ddxideltaFxi},
\begin{align}
\frac{d}{d\xi}\delta C(\xi)={}&\tfrac{1}{2}\,{\rm Tr}\,\bigl(\tilde{S}S^\dagger\Xi(-\xi)\Xi'(\xi)\tilde{S}S^\dagger-\Xi'(\xi)\tilde{S}S^\dagger\bigr)\nonumber\\
={}&\tfrac{1}{2}\,{\rm Tr}\,\bigl([i\sigma_y({\cal F}-{\cal F}_c)(\cos\xi-1)+\sin\xi](\tilde{S}S^\dagger)^2\nonumber\\
&-[({\cal F}-{\cal F}_c)\sin\xi+i\sigma_y\cos\xi]\tilde{S}S^\dagger\bigr).
\end{align}

Using again the identity \eqref{keyidentity}, as well the fact that $\tilde{S}S^\dagger$ commutes with $\sigma_y$, we arrive at
\begin{align}
\frac{d}{d\xi}\delta C(\xi)={}&-\tfrac{1}{2}i\,{\rm Tr}\,\sigma_y\tilde{S}S^\dagger\nonumber\\
&+\tfrac{1}{2}i(1-\cos\xi)\,{\rm Tr}\,\sigma_y[\tilde{S}S^\dagger-(\tilde{S}S^\dagger)^3].
\end{align}
The first $\xi$-independent trace is $i$ times the average charge \eqref{firsttwocumulantsa}. The second trace is an even function of $\xi$, meaning that it produces only odd cumulants --- all even cumulants vanish.

\subsection{Odd cumulants}

We have not yet used specific properties of the phase profile $\eta(t)$ that characterizes the adiabatic scattering matrix $S_0(t)=e^{i\eta(t)\sigma_y}$. In the previous subsection we showed that the even cumulants vanish for any $\eta(t)$. In this subsection we show that the odd cumulants vanish if the net increment
\begin{equation}
\Delta\eta=\int_{-\infty}^\infty \eta'(t)\,dt
\end{equation}
is an integer multiple of $\pi$, which is the case considered here --- but not more generally.

The calculation of the odd cumulants is based on the equation
\begin{equation}
{\rm Tr}\,\sigma_y(\tilde{S}S^\dagger)^3=\frac{1}{\pi}(\Delta\eta-\sin\Delta\eta),\label{keyidentity2}
\end{equation}
derived in App.\ \ref{app_identitiesB}.
Since
\begin{equation}
{\rm Tr}\,\sigma_y\tilde{S}S^\dagger=\frac{1}{\pi}\Delta\eta,
\end{equation}
in view of Eq.\ \eqref{tildeSresult}, we find
\begin{align}
\frac{d}{d\xi}\delta C(\xi)={}&-\frac{1}{2\pi}i\Delta\eta+\frac{1}{2\pi}i(1-\cos\xi)\sin\Delta\eta,
\end{align}
which gives the nonequilibrium contribution to the cumulants,
\begin{equation}
\delta\langle\!\langle Q^p\rangle\!\rangle=-\frac{1}{2\pi}\times\begin{cases}
\Delta\eta&\text{for}\;\;p=1,\\
0&\text{for even}\;\;p\geq 2,\\
\sin\Delta\eta&\text{for odd}\;\;p\geq 3.
\end{cases}\label{allcumulants}
\end{equation}

\section{How to reduce the equilibrium noise}
\label{sec_reduce}

The equilibrium noise \eqref{eqnoisefinal} persists at zero temperature \cite{Lev96},
\begin{equation}
\lim_{T\rightarrow 0}\langle\!\langle Q^2\rangle\!\rangle_{\rm eq}=\frac{e^2}{2\pi^2}\ln(t_{\rm det}/\epsilon),\;\;t_{\rm det}\gg\epsilon.\label{eqnoisezeroT}
\end{equation}
Would it be possible to suppress this noise and observe only the contribution from the nonequilibrium charge transfer?

The established procedure for integer charge pumping \cite{Lev96,Iva97} is to repeat the pumping process periodically for a large number $N$ of cycles. The variance of the transferred charge then contains a contribution $\propto \ln N$ from the equilibrium fluctuations, and a contribution $\propto N$ from the pumping process, so the equilibrium noise drops out of the noise per cycle $\lim_{N\rightarrow\infty}N^{-1}\,{\rm Var}\,Q$.

This procedure fails in our case of half-integer charge transfer, because the vortex injection process is not periodic: the $2\pi$ increment of the superconducting phase flips the sign of single-electron wave functions. After a $4\pi$ increment the periodicity is restored, but then we are back to an integer charge transfer. Choosing the $4\pi$ increment as a periodically repeated cycle and counting only the charge from the first $2\pi$ increment does not provide a workaround, because in that case the nonequilibrium noise will scale $\propto N$ rather than $\propto \ln N$ --- for the $\ln N$ scaling it is essential that the counting process is not interrupted within a cycle.

The result \eqref{eqnoisezeroT} holds for a step function counting window, where charge is counted with unit efficiency in the interval $-t_{\rm det}/2<t<t_{\rm det}/2$. Smoothing of the abrupt switch-on/switch-off over a time scale $t_{\rm smooth}$, for example, by convolution of the step function with a Gaussian \cite{Iva16}, has the effect of replacing the inverse bandwidth $\epsilon$ by $t_{\rm smooth}$,
\begin{equation}
\langle\!\langle Q^2\rangle\!\rangle_{\rm eq}=\frac{e^2}{2\pi^2}\ln(t_{\rm det}/t_{\rm smooth}),\;\;t_{\rm det}\gg t_{\rm smooth}\gg\epsilon.\label{varWtsmooth}
\end{equation}
Hence the logarithmically increasing equilibrium noise is not removed by a smoothed counting window. To reduce $\langle\!\langle Q^2\rangle\!\rangle_{\rm eq}$ we need a \textit{heavy-tailed} profile, as we now discuss.

The counting profile $W(t)$ replaces $\int_{-t_{\rm det}/2}^{t_{\rm det}/2}dt$ by $\int_{-\infty}^\infty W(t)dt$ in the expression \eqref{Qaveragefinal} for the average charge,
\begin{equation}
\langle Q\rangle=-\frac{e}{2\pi}\int_{-\infty}^{\infty} W(t)dt\,\eta'(t),\label{QaverageWt}
\end{equation}
and in the expression \eqref{varQeqFttprime} for the equilibrium noise,
\begin{align}
&\langle\!\langle Q^2\rangle\!\rangle_{\rm eq}=e^2\int_{-\infty}^\infty W(t)dt\int_{-\infty}^\infty W(t')dt'\,{\cal F}(t,t'){\cal F}_c(t',t)\nonumber\\
&\qquad=e^2\int_{-\infty}^\infty  \frac{dE}{2\pi}\int_{-\infty}^\infty \frac{dE'}{2\pi}\,|W(E'-E)|^2f(E)f(-E').
\end{align}
We have defined the Fourier transform $W(E)=\int_{-\infty}^\infty dt\,e^{iEt}W(t)$. At $T=0$ the Fermi function $f(E)$ becomes a step function, hence
\begin{equation}
\lim_{T\rightarrow 0}\langle\!\langle Q^2\rangle\!\rangle_{\rm eq}=\frac{e^2}{4\pi^2}\int_0^\infty dE\,E|W(E)|^2.\label{varWTiszero}
\end{equation}

The Fourier transform of a heavy-tailed $W(t)$ is a stretched exponential \cite{note2},
\begin{equation}
W(E)=\frac{\pi t_W}{\Gamma(1+1/\alpha)}e^{-|Et_W|^\alpha},\;\;0<\alpha\leq 2.\label{Westable}
\end{equation}
The normalization is adjusted such that if we send the time constant $t_W$ to infinity at fixed measurement time $t$, the counting efficiency $W(t)\rightarrow 1$. This ensures that $\lim_{t_W\rightarrow\infty}\langle Q\rangle=\pm e/2$. The zero-temperature equilibrium noise \eqref{varWTiszero} is $t_W$-independent, equal to 
\begin{equation}
\langle\!\langle Q^2\rangle\!\rangle_{\rm eq}=\frac{e^2\Gamma \left(\tfrac{1}{2}+\frac{1}{\alpha}\right)}{8 \sqrt{\pi } \Gamma \left(1+\frac{1}{\alpha}\right)}< \frac{e^2\sqrt{\alpha}}{8\sqrt{\pi}}.
\end{equation}

\begin{figure}[tb]
\centerline{\includegraphics[width=0.9\linewidth]{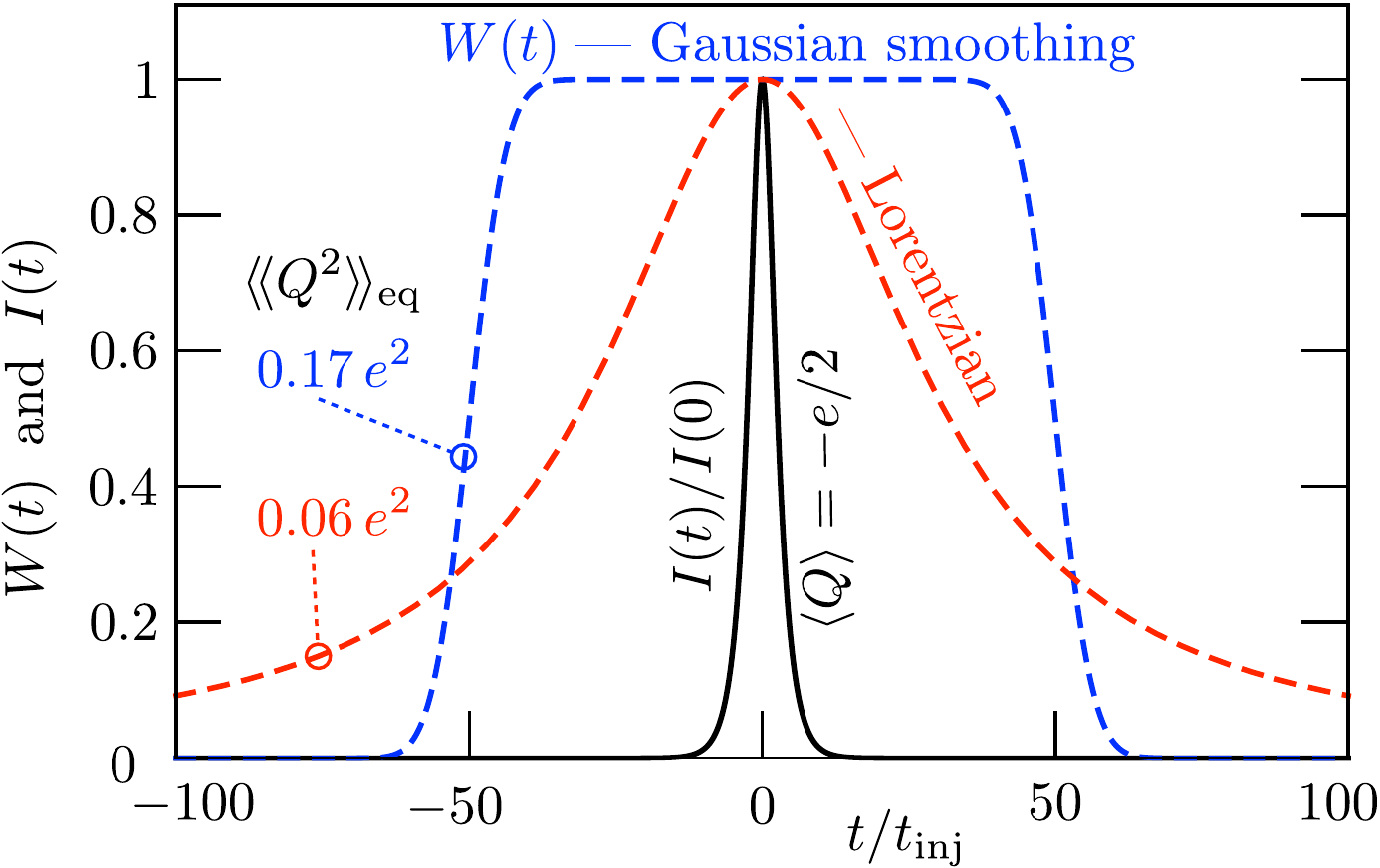}}
\caption{Comparison of the effect of two charge counting profiles on the zero-temperature equilibrium charge noise. The solid curve shows the current $I(t)=(e/2\pi)\eta'(t)$, with $\eta(t)$ given by Eq.\ \eqref{etawidejunction}. A charge $-e/2$ is transferred on a time scale $t_{\rm inj} $. The blue dashed curve is a step function counting window (width $t_{\rm det}=100\,t_{\rm inj}$), smoothed by convoluting with a Gaussian (width $t_{\rm smooth}=5\,t_{\rm inj}$). The corresponding charge variance \cite{note3} is $0.166\,e^2$. The red dashed curve is a Lorentzian counting window, $W(t)=[1+(\pi t/t_W)^2]^{-1}$ with $t_{\rm W}=100\,t_{\rm inj}$. The variance is reduced to $e^2/16$.
}
\label{fig_smooth}
\end{figure}

The parameter $\alpha$ need not become very small for a substantially reduced noise, the Lorentzian case $\alpha=1\Rightarrow W(t)=[1+(\pi t/t_W)^2]^{-1}$ has $\langle\!\langle Q^2\rangle\!\rangle_{\rm eq}=e^2/16$. In Fig.\ \ref{fig_smooth} we compare this heavy-tailed counting profile with the smoothed step function. In this case the noise reduction is about a factor of 3, by lowering $\alpha$ and raising $t_{\rm W}$ the noise can be reduced further.

The reduction of the equilibrium noise $\langle\!\langle Q^2\rangle\!\rangle_{\rm eq}$ leaves the nonequilibrium noise unaffected: $\delta\langle\!\langle Q^2\rangle\!\rangle\rightarrow 0$ for $t_W/ t_{\rm inj}\rightarrow\infty$. We can therefore reach a situation of nearly noiseless transfer of a half-integer charge.

\section{Conclusion}
\label{sec_conclude}

In conclusion, we have given an affirmative answer to the question whether the half-integer charge transferred into a normal metal by the fusion of a pair of Majorana edge modes can be considered a sharp observable. The integer value constraint of a charge counting measurement is avoided by the equilibrium noise, which persists at zero temperature in a gapless system. We have shown that the nonequilibrium charge transfer does not produce any excess noise --- all cumulants of transferred charge retain their equilibrium value.

Our analysis is based on the fermionic scattering approach to the cumulant generating function pioneered by Levitov, Lee, and Lesovik \cite{Lev96}. The same authors also proposed an alternative approach based on the technique of bosonisation. This approach has been applied to the Majorana edge mode problem in Ref.\ \onlinecite{Ada20}, and in App.\ \ref{sec_boson} we show that it also gives noiseless $e/2$ charge transfer.

The equilibrium noise can be reduced by measuring the charge with a heavy-tailed detection efficiency. A specific example of a Lorentzian detection profile gives a variance of $e^2/16$ --- well below the value $e^2/4$ that would follow if the $e/2$ charge transfer would follow a binomial statistics. A measurement of such a nearly noiseless half-integer charge transfer would be a milestone in the field of charge fractionalization, by extending the familiar physics of localized zero-modes \cite{Jac76,Su79,Nie86,Hou07,Ben19,Kiv82,Bel82,Jac83} to mobile edge modes. External sources of charge noise, from low-energy quasiparticles that may be present in the superconductor, should be suppressed for such an experiment to succeed.

\begin{figure}[tb]
\centerline{\includegraphics[width=0.9\linewidth]{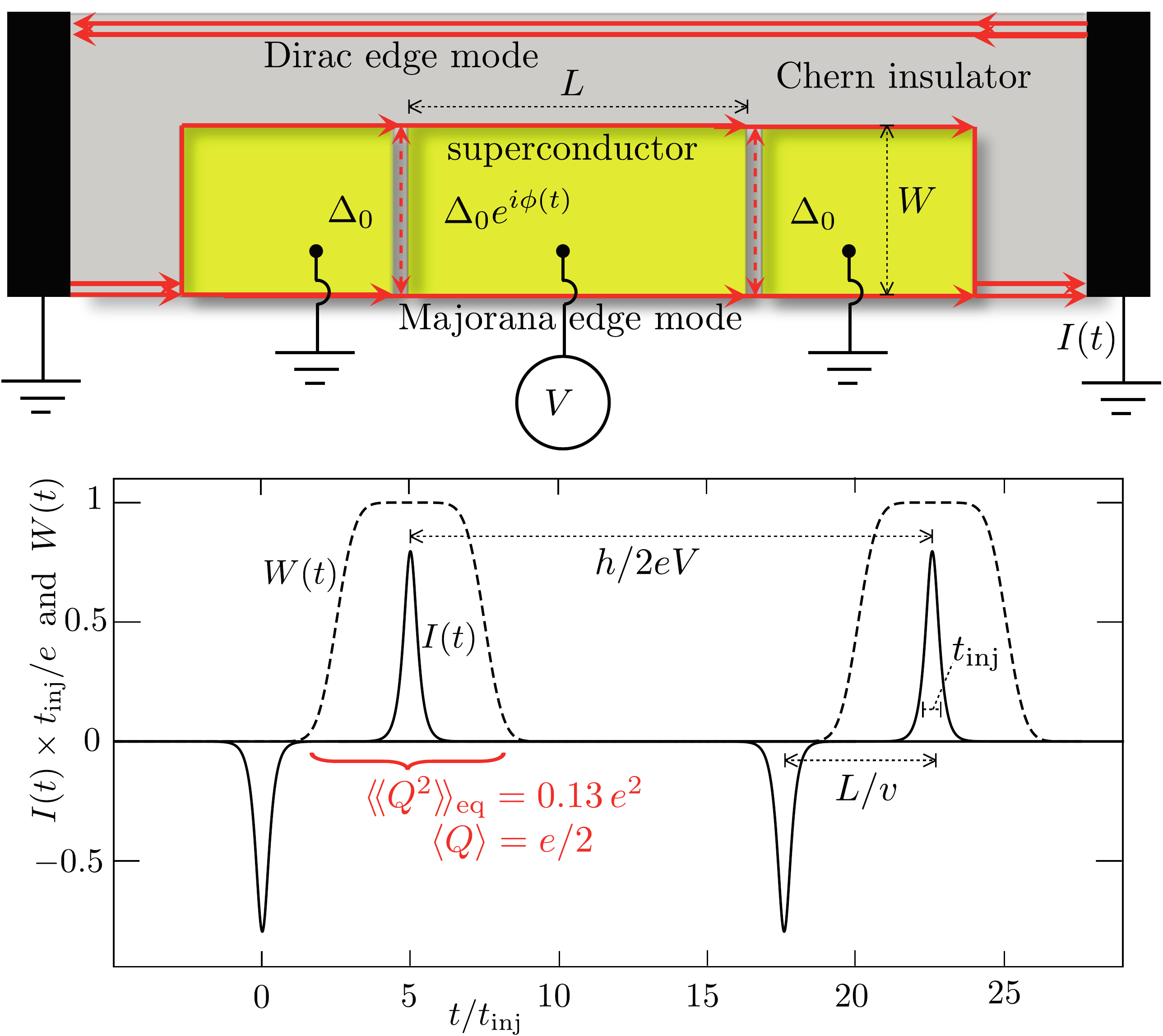}}
\caption{Top panel: Geometry from Ref.\ \onlinecite{Bee19a} to inject Majorana edge vortices in a Chern insulator/superconductor heterostructure. Alternatively to Fig.\ \ref{fig_layout}, the edge modes are excited by a constant voltage $V$ rather than a time-dependent flux. A pair of current pulses $I(t)$ of opposite sign, of width $t_{\rm inj}$ and spaced in time by $L/v$, is injected with period $h/2eV$. A detection window $W(t)$ selects one of the two pulses, measuring an average charge $\langle Q\rangle=e/2$ per period. The profile in the lower panel shows current pulses according to Eqs.\ \eqref{etawidejunction} and \eqref{Qaveragefinal}, and a detection window with Gaussian smoothing ($t_{\rm det}=5t_{\rm inj}$, $t_{\rm smooth}=0.5\,t_{\rm inj}$), for which the equilibrium charge noise at zero temperature is $\langle\!\langle Q^2\rangle\!\rangle_{\rm eq}=0.13\,e^2$ per period \cite{note3}. The theory presented in this work predicts that the half-integer charge transfer produces no excess noise, so this would be the entire noise measured --- well below the variance of $\frac{1}{4}e^2$ expected for binomial transfer statistics.
}
\label{fig_layout2}
\end{figure}

A measurement configuration using a flux-biased Josephson junction on the surface of a topological insulator is shown in Fig.\ \ref{fig_layout}. Fig.\ \ref{fig_layout2} shows an alternative configuration using a voltage-biased junction on a Chern insulator \cite{She20}. In that setup a periodic train of charges $\pm e/2$ is injected by a {\sc dc} voltage $V$. The width of the charge pulses is given by Eq.\ \eqref{tphidef},
\begin{equation}
t_{\rm inj}=\frac{\xi_0}{W}\frac{\hbar}{2eV},
\end{equation}
it is much less than the spacing $h/2eV$ of the pulses for $\xi_0\ll W$. 

This is a key difference between charge pumping by a Josephson junction in a superconductor and by a tunnel junction in a normal metal: In the normal metal a narrow voltage pulse is needed in order to obtain well-separated charge pulses \cite{Lev96}. Half-integer charge injection by a tunnel junction requires fine tuning to $h/2e$ of the area $\int V(t)dt$ under the voltage pulse. The spectral properties of such an excitation of the Fermi sea have been studied \cite{Mos16,Yue20}. We expect the excess noise for chiral propagation to vanish in that case as well.

\acknowledgments

This project has received funding from the Netherlands Organization for Scientific Research (NWO/OCW) and from the European Research Council (ERC) under the European Union's Horizon 2020 research and innovation programme. FH acknowledges support by the Deutsche Forschungsgemeinschaft (DFG, German Research Foundation) under Germany's Excellence Strategy -- Cluster of Excellence Matter and Light for Quantum Computing (ML4Q) EXC 2004/1 -- 390534769.

\appendix

\section{Effect on the excess charge noise of a nonzero relative delay}
\label{app_delay}

In the main text we assumed that the path length from Josephson junction to metal contact was the same along upper and lower edge. Here we relax that assumption, and allow for a path length difference $\delta L$, corresponding to a relative time delay $\delta t=\delta L/v$. We take the limit of infinitely large detection time $t_{\rm det}$. The results then depend on the ratio of $\delta t$ and the injection time $t_{\rm inj}$, as well as on the product $T\delta t$.

The scattering matrix \eqref{Swithdelay} is anchored to the Fermi level via the commutator $\tilde{S}=[{\cal F},S]$ with the Fermi function, which gives  
\begin{align}
  &\tilde{S}(t,t') =
  -f(t-t')
  \begin{pmatrix}
    1 & 0 \\
    0 & 0
  \end{pmatrix}
  \bigl(S_0(t)-S_0(t')\bigr)\nonumber\\
  &\qquad
  - f(t-t'+\delta t)
  \begin{pmatrix}
    0 & 0 \\
    0 & 1
  \end{pmatrix}
  \bigl(S_0(t+\delta t)-S_0(t')\bigr),
\end{align}
with
\begin{equation}
{\cal F}(t,t')=f(t-t'),\;\;f(t) = \frac{{i} T}{2 \sinh[\pi T (t+i\epsilon)]}.
\end{equation}
We also need
\begin{align}
&[S^{\dagger}\sigma_y\tilde{S}](t,t')=\nonumber\\
&=f(t-t'+\delta t)S_0^\dagger(t)\begin{pmatrix}
0&i\\
0&0
\end{pmatrix}[S_0(t+\delta t)-S_0(t')]\nonumber\\
&-f(t-t'-\delta t)S_0^\dagger(t)\begin{pmatrix}
0&0\\
i&0
\end{pmatrix}[S_0(t-\delta t)-S_0(t')].\label{appSdaggersigmayS}
\end{align}
The kernel $S^\dagger({\cal F}-{\cal F}_c)\tilde{S}$ is the same as for $\delta t=0$,
\begin{align}
&[S^\dagger({\cal F}-{\cal F}_c)\tilde{S}](t_1,t_2)=\nonumber\\
&\qquad=\int dt\, f(t-t_2)g(t_1-t)S_0^\dagger(t_1)[S_0(t_2)-S_0(t)],\nonumber\\
&g(t)={\cal P}\frac{iT}{\sinh(\pi T t)}.
\end{align}

Substitution of Eq.\ \eqref{appSdaggersigmayS} into the general expression \eqref{firsttwocumulantsa} for the average charge, with $S_0=e^{i\eta(t)\sigma_y}$, gives
\begin{equation}
    \langle Q\rangle = -\frac{e}{2}T\int dt\,
    \frac{\sin[\eta(t+\delta t) - \eta(t)]}{ \sinh(\pi T \delta t)}.\label{Qdeltattemp}
\end{equation}
In the limit $\delta t\rightarrow 0$ we recover Eq.\ \eqref{Qaveragefinal}.

For the excess noise \eqref{firsttwocumulantsc} we need
\begin{widetext}
\begin{align}
{\rm Tr}\,[S^\dagger({\cal F}-{\cal F}_c) \tilde{S}]={}& 2 T^2 \int {d} t \int {d} t' \:
    \frac{\sin^2 [\tfrac{1}{2}\eta(t)-\tfrac{1}{2}\eta(t')]}{\sinh^2[\pi T(t-t')]},\\
{\rm Tr}\,[S^\dagger \sigma_y \tilde{S} S^\dagger \sigma_y \tilde{S}]
  ={}& 2 T^2 \int {d} t \int {d} t' \,
  \frac{\sin^2  [\tfrac{1}{2}\eta(t)-\tfrac{1}{2}\eta(t')]
    \sin^2  [\tfrac{1}{2}\eta(t+\delta t)-\tfrac{1}{2}\eta(t'+\delta t)]}
  {\sinh^2[\pi T(t-t')]}\nonumber\\
  &  + \tfrac{1}{2}T^2\int {d} t \int {d} t' \,
  \frac{\sin[\eta(t+\delta t)-\eta(t')]
    \sin [\eta(t)-\eta(t' + \delta t)]}
  {\sinh[\pi T(t-t'+\delta t)] \sinh[\pi T(t-t'+\delta t)]}.\end{align}
Combining results, we arrive at the delay-time dependent excess noise
\begin{align}
 \delta\langle\!\langle Q\rangle\!\rangle
  ={}& T^2 \int {d} t \int {d} t' \,
  \frac{\sin^2  [\tfrac{1}{2}\eta(t)-\tfrac{1}{2}\eta(t')]
    \cos^2  [\tfrac{1}{2}\eta(t+\delta t)-\tfrac{1}{2}\eta(t'+\delta t)]}
  {\sinh^2[\pi T(t-t')]}\nonumber
  \\
  &-T^2 \int {d} t \int {d} t' \,
  \frac{\sin[\eta(t+\delta t)-\eta(t')]
    \sin [\eta(t)-\eta(t' + \delta t)]}
  {4\sinh[\pi T(t-t'+\delta t)]\sinh[\pi T(t-t'-\delta t)]}.\label{deltaQtemp}
\end{align}
\end{widetext}
For $\delta t=0$ the two integrands cancel, as obtained in the main text. Fig.\ \ref{fig_noise} shows a plot for the case $W\gg \xi_0$ of a wide Josephson junction, when $\eta(t)$ is given by Eq.\ \eqref{etawidejunction}.

\section{Details of the calculation of higher cumulants}
\label{app_identities}

\subsection{Derivation of Eq.\ \eqref{keyidentity}}
\label{app_identitiesA}

To derive the identity \eqref{keyidentity}, needed for the calculation of the even cumulants, we consider the operator product $({\cal F}-{\cal F}_c)\tilde{S}S^\dagger$ in the $T=0$ limit. Upon substitution of Eqs.\ \eqref{barFminF} and \eqref{tildeSresult} we have
\begin{equation}
[({\cal F}-{\cal F}_c)\tilde{S}S^\dagger](t_1,t_2)=\dashint_{-\infty}^\infty dt\,\frac{S_0(t)S_0^\dagger(t_2)-1}{2\pi^2(t_1-t) (t-t_2)},\label{longkernel}
\end{equation}
where $\dashint$ denotes the Cauchy principal value integral. Because
\begin{equation}
\dashint_{-\infty}^\infty \frac{dt}{(t_1-t)(t-t_2)}=0,\label{PVidentity}
\end{equation}
for any $t_1,t_2$, including $t_1=t_2$ (see App.\ C of Ref.\ \onlinecite{Dav90}), this reduces to
\begin{equation}
[({\cal F}-{\cal F}_c)\tilde{S}S^\dagger](t_1,t_2)=\dashint_{-\infty}^\infty dt\,\frac{S_0(t)S_0^\dagger(t_2)}{2\pi^2(t_1-t) (t-t_2)}.
\end{equation}
Similarly,
\begin{equation}
[\tilde{S}S^\dagger({\cal F}-{\cal F}_c)](t_1,t_2)=\dashint_{-\infty}^\infty dt\,\frac{S_0(t_1)S_0^\dagger(t)}{2\pi^2(t_1-t) (t-t_2)}.\label{longkernel2}
\end{equation}

We next evaluate, still in the $T\rightarrow 0$ limit,
\begin{widetext}
\begin{align}
[(\tilde{S}S^\dagger)^2](t_1,t_2)={}&\int_{-\infty}^\infty dt\,\frac{(S_0(t_1)S_0^\dagger(t)-1)(S_0(t)S_0^\dagger(t_2)-1)}{-4\pi^2(t_1-t)(t-t_2)}\nonumber\\
={}&\dashint_{-\infty}^\infty dt\,\frac{S_0(t_1)S_0^\dagger(t)+S_0(t)S_0^\dagger(t_2)}{4\pi^2(t_1-t)(t-t_2)}-(S_0(t_1)S_0^\dagger(t_2)+1)\dashint_{-\infty}^\infty \frac{dt}{4\pi^2(t_1-t)(t-t_2)}\nonumber\\
={}&\dashint_{-\infty}^\infty dt\,\frac{S_0(t_1)S_0^\dagger(t)+S_0(t)S_0^\dagger(t_2)}{4\pi^2(t_1-t)(t-t_2)}.
\end{align}
\end{widetext}
In the last equation we have again used Eq.\ \eqref{PVidentity}. Comparison with Eqs.\ \eqref{longkernel} and \eqref{longkernel2} produces the identity \eqref{keyidentity} in the main text.

\subsection{Derivation of Eq.\ \eqref{keyidentity2}}
\label{app_identitiesB}

We give a proof of the identity \eqref{keyidentity2}, needed for a calculation of the odd cumulants. We evaluate ${\rm Tr}\,\sigma_y(\tilde{S}S^\dagger)^3$ at $T=0$, by substituting Eq.\ \eqref{tildeSresult} with $S(t)=e^{i\sigma_y\eta(t)}$,
\begin{widetext}
\begin{align}
{\rm Tr}\,\sigma_y(\tilde{S}S^\dagger)^3={}&\frac{1}{(2\pi i)^3}\int_{-\infty}^\infty dt_1\int_{-\infty}^\infty dt_2\,\int_{-\infty}^\infty dt_3\,{\rm Tr}\,\frac{S_0(t_1)S^\dagger_0(t_2)-1}{t_1-t_2}\frac{S_0(t_2)S^\dagger_0(t_3)-1}{t_2-t_3}\frac{S_0(t_3)S^\dagger_0(t_1)-1}{t_3-t_1}\nonumber\\
={}&\frac{4i}{(2\pi i)^3}\int_{-\infty}^\infty dt_1\int_{-\infty}^\infty dt_2\,\int_{-\infty}^\infty dt_3\,\frac{\sin[\eta(t_1)-\eta(t_2)]+ \sin[\eta(t_2)-\eta(t_3)]+\sin[\eta(t_3)-\eta(t_1)]}{(t_1-t_2)(t_2-t_3)(t_3-t_1)}.\label{trm3threesines}
\end{align}
\end{widetext}

The full integrand \eqref{trm3threesines}, including all three sines in the numerator, is nonsingular, but we will be breaking it up into three separate sine contributions that are individually singular. We regularize the singularities in two ways: Firstly, we avoid the poles at $t=t'$ by inserting a positive infinitesimal $\epsilon$ in each denominator, $t-t'\mapsto t-t'+i\epsilon$. Secondly, to ensure that individual contributions $\propto\sin[\eta(t)-\eta(t')]$ vanish for $|t-t'|\rightarrow\infty$ we substract
\begin{equation}
L(t,t')=[\theta(t)\theta(-t')-\theta(t')\theta(-t)] \sin\Delta\eta,
\end{equation}
with $\Delta\eta=\eta(\infty)-\eta(-\infty)$. This has no effect on the integrand in Eq.\ \eqref{trm3threesines}, since
\begin{equation}
L(t_1,t_2)+L(t_2,t_3)+L(t_3,t_1)=0.
\end{equation}
The integral \eqref{trm3threesines} then has three identical contributions,
\begin{widetext}
\begin{align}
{\rm Tr}\,\sigma_y(\tilde{S}S^\dagger)^3=\frac{12i}{(2\pi i)^3}\int_{-\infty}^\infty dt_1\int_{-\infty}^\infty dt_2\,\int_{-\infty}^\infty dt_3\,\frac{\sin[\eta(t_1)-\eta(t_2)]-L(t_1,t_2)}{(t_1-t_2+i\epsilon)(t_2-t_3+i\epsilon)(t_3-t_1+i\epsilon)}.\label{trm3threesines2}
\end{align}
 
The integral over $t_3$ is carried out by closing the contour in the upper half of the complex plane, picking up the pole at $t_2+i\epsilon$,
\begin{equation}
{\rm Tr}\,\sigma_y(\tilde{S}S^\dagger)^3=\frac{24\pi}{(2\pi i)^3}\int_{-\infty}^\infty dt_1\int_{-\infty}^\infty dt_2\,\frac{\sin[\eta(t_1)-\eta(t_2)]-L(t_1,t_2)}{(t_1-t_2+i\epsilon)(t_2-t_1+2i\epsilon)}.
\end{equation}
We thus need to evaluate the integral
\begin{equation}
{\rm Tr}\,\sigma_y(\tilde{S}S^\dagger)^3 =\frac{3}{\pi^2}\int_{-\infty}^\infty dt\int_{-\infty}^\infty dt'\,\frac{\sin[\eta(t)-\eta(t')]-L(t,t')}{t-t'}u(t-t'),\;\;u(t)=\frac{\epsilon t^2}{(t^2+\epsilon^2)(t^2+4\epsilon^2)},\label{trm3singlesine}
\end{equation}
discarding terms that are odd under interchange $t\leftrightarrow t'$ and vanish upon integration.
\end{widetext}

In the limit $\epsilon\rightarrow 0$, the function $u(t)$ becomes a delta function,
\begin{equation}
\lim_{\epsilon\rightarrow 0}u(t)=\tfrac{1}{3}\pi\delta(t).
\end{equation}
Since $\lim_{t\rightarrow t'} L(t,t')/(t-t')=\delta(t-t')\sin\Delta\eta$, we arrive at
\begin{align}
{\rm Tr}\,\sigma_y(\tilde{S}S^\dagger)^3&=\frac{1}{\pi}\int_{-\infty}^\infty dt\,[\eta'(t)-\delta(t)\sin\Delta\eta]\nonumber\\
&=\frac{1}{\pi}(\Delta\eta-\sin\Delta\eta),
\end{align}
which is Eq.\ \eqref{keyidentity2} in the main text.

\section{Cumulant generating function via bosonisation}
\label{sec_boson}

We give an alternative calculation of the current fluctuations using the bosonisation approach to counting statistics \cite{Lev96}. We summarize equations from Ref.\ \onlinecite{Ada20}, where this approach was applied to Majorana edge modes. 

We transform to a coordinate frame that moves along the edge with velocity $v\equiv 1$, so the independent space and time variables are $s=x-t$ and $\tau=t+x$. The regularized density operator of the chiral mode is a Hermitian bosonic field $\hat\rho(s)$, with commutator
\begin{equation}
[\hat{\rho}(s),\hat{\rho}(s')]=\frac{i}{2\pi}\frac{\partial}{\partial s}\delta(s-s').\label{KacMoody}
\end{equation}
A many-body state evolves according to $|\tau\rangle=\hat{\cal S}(\tau)|0\rangle$, with unitary scattering operator ${\cal S}(\tau)$. The corresponding evolution of the density operator is determined by 
\begin{equation}
\hat{\cal S}^\dagger(\tau)\hat{\rho}(s)\hat{\cal S}(\tau)=\hat{\rho}(s)+\frac{1}{2\pi}\frac{\partial}{\partial s}\Lambda(s,\tau).\label{SdaggerrhosS}
\end{equation}
The field $\Lambda(s,\tau)$ is related to the phase profile $\eta(t)$ in the main text by
\begin{equation}
\Lambda(s,\tau)=\eta(s+\tau)\eta(-s).
\end{equation}

We now calculate the full counting statistics of the transferred charge $\hat{Q}=\int W(s)\hat{\rho}(s)\,ds$, with detection profile $W(s)$. The cumulant generating function is
\begin{align}
C(\xi)&=\ln\langle\tau|e^{i\xi\hat{Q}}|\tau\rangle\nonumber\\
&=\ln\left\langle 0\left|\exp\left(i\xi\int W(s)ds\,\hat{\cal S}^\dagger(\tau) \hat{\rho}(s)\hat{\cal S}(\tau)\right)\right|0\right\rangle\nonumber\\
&=\frac{i\xi}{2\pi}\int W(s)ds\frac{\partial}{\partial s}\Lambda(s,\tau)+C_{\rm eq}(\xi),\label{Zxis}
\end{align}
where $C_{\rm eq}(\xi)=\ln\langle 0|e^{i\xi\hat{Q}}|0\rangle$ gives the equilibrium fluctuations. Because $\delta C(\xi)=C(\xi)-C_{\rm eq}(\xi)$ is linear in $\xi$, the nonequilibrium current is noiseless.

It is satisfying to see how the noiseless $e/2$ charge transfer is an immediate consequence of the evolution equation \eqref{SdaggerrhosS} of the bosonic field. In the main text we have followed the more computationally intensive fermionic approach, because of a difficulty which we have encountered in the general case of arbitrary fractional charge transfer: from Eq.\ \eqref{Zxis} we would conclude that $ C(\xi)-C_{\rm eq}(\xi)$ is linear in $\xi$ for \textit{any} phase profile $\eta(t)$, irrespective of the increment $\Delta\eta=\eta(\infty)-\eta(-\infty)$. This disagrees with the calculation using the fermionic scattering approach, which finds odd cumulants $\propto\sin\Delta\eta$ in Eq.\ \eqref{allcumulants}. 

Difficulties with the bosonisation approach to full counting statistics have been noted before \cite{Iva16}, but those addressed differences between smooth and abrupt detection profiles $W(s)$. The difficulty signaled here seems to be of a different nature and calls for further investigation.


\begin{thebibliography}{99}
\bibitem{note1} The need for a regularization of the determinant \eqref{Fxidet} is immediately obvious in the adiabatic approximation, when $S$ commutes with $\sigma_y$ and would drop out of the cumulant generating function. The adiabatic approximation holds only near the Fermi level, while Eq.\ \eqref{Fxidet} contains contributions from deep in the Fermi sea. By anchoring the scattering matrix to the Fermi level this complication is avoided.
\bibitem{note2} The restriction to $\alpha\leq 2$ in Eq.\ \eqref{Westable} ensures that $W(t)>0$ for all $t$. This counting profile is known in probability theory as a ``stable distribution'', see the Wikipedia entry for a summary of its properties.
\end{thebibliography}
\end{document}